\def\year{2022}\relax
\documentclass[letterpaper]{article} % DO NOT CHANGE THIS
\usepackage{aaai22}  % DO NOT CHANGE THIS
\usepackage{times}  % DO NOT CHANGE THIS
\usepackage{helvet}  % DO NOT CHANGE THIS
\usepackage{courier}  % DO NOT CHANGE THIS
\usepackage[hyphens]{url}  % DO NOT CHANGE THIS
\usepackage{graphicx} % DO NOT CHANGE THIS
\usepackage{natbib}  % DO NOT CHANGE THIS AND DO NOT ADD ANY OPTIONS TO IT
\usepackage{caption} % DO NOT CHANGE THIS AND DO NOT ADD ANY OPTIONS TO IT
\DeclareCaptionStyle{ruled}{labelfont=normalfont,labelsep=colon,strut=off} % DO NOT CHANGE THIS
\usepackage{algorithm}
\usepackage{algorithmic}
\usepackage{amsmath}
\usepackage{newfloat}
\usepackage{listings}

\usepackage{makecell}
\usepackage{multirow}
\usepackage{rotating}
\usepackage{threeparttable} % table remark
\usepackage{amssymb}
\usepackage{amsthm}
\theoremstyle{definition}
\newtheorem{definition}{Definition}

\usepackage[algo2e]{algorithm2e} 
\SetKwInput{KwInput}{Input}                % Set the Input
\SetKwInput{KwOutput}{Output}              % set the Output

\usepackage{booktabs}
\usepackage{arydshln}

\floatstyle{ruled}
\newfloat{listing}{tb}{lst}{}
\floatname{listing}{Listing}
\title{BlockGC: A Joint Learning Framework for Account Identity Inference \\on Blockchain with Graph Contrast}
\author {
    Jiajun Zhou\textsuperscript{\rm 1},
    Chenkai Hu\textsuperscript{\rm 1},
    Shenbo Gong\textsuperscript{\rm 1},
    Jiaying Xu\textsuperscript{\rm 1},
	Jie Shen\textsuperscript{\rm 1},
    Qi Xuan\textsuperscript{\rm 1} \thanks{Qi Xuan is the corresponding author.}
}
\affiliations{
    \textsuperscript{\rm 1}Zhejiang University of Technology, Hangzhou, China, \\ 
	\{jjzhou, shenj, xuanqi\}@zjut.edu.cn, \{ckhu0122, jshmhsb, JiayingXu1279\}@gmail.com
}
\usepackage{bibentry}
\begin{document}

\maketitle

\begin{abstract}
	Blockchain technology has the characteristics of decentralization, traceability and tamper proof, which creates a reliable decentralized transaction mode, further accelerating the development of the blockchain platforms.
	% which not only lays a firm foundation of trust but also creates a reliable decentralized transaction mode, further accelerating the development of the blockchain platforms.
	However, with the popularization of various financial applications, security problems caused by blockchain digital assets, such as money laundering, illegal fundraising and phishing fraud, are constantly on the rise.
	Therefore, financial security has become an important issue in the blockchain ecosystem, and identifying the types of accounts in blockchain (e.g. miners, phishing accounts, Ponzi contracts, etc.) is of great significance in risk assessment and market supervision.
	% For identity inference for accounts in blockchain, widely used methods concentrate on manual feature engineering and graph embedding, suffering from several shortcomings. 
	In this paper, we construct an account interaction graph using raw blockchain data in a graph perspective, and proposes a joint learning framework for account identity inference on blockchain with graph contrast.
	We first capture transaction feature and correlation feature from interaction graph, and then perform sampling and data augmentation to generate multiple views for account subgraphs, finally jointly train the subgraph contrast and account classification task.
	Extensive experiments on Ethereum datasets show that our method achieves significant advantages in account identity inference task in terms of classification performance, scalability and generalization.

\end{abstract}

\section{Introduction}\label{sec: introduction}
The past few years have witnessed the application of blockchain technology in new technological and industrial revolutions, such as cryptocurrency, supply chain management, financial services, etc.
As a distributed data storage technology, blockchain is decentralized, traceable and tamper-proof, which guarantees the fidelity and security of a record of data and generates trust without the need for a trusted third party.
Benefitting from these characteristics, blockchain has attracted considerable attention, and are best known for its crucial role in cryptocurrency systems, such as Bitcoin and Ethereum.
% According to statistics from market analysis sites such as CoinMarketCap, as of August 2021, the existing types of cryptocurrencies have reached about 11000, with a total market value of up to 1.9 trillion dollars.

However, while the anonymity mechanism of the blockchain achieves decentralization, it also magnifies the security risk of cryptoasset.
Blockchain is most widely used in financial transaction, making it a tempting target for hackers and other cybercriminals.
At present, the weak supervision of blockchain platforms has led to endless illegal criminal activities such as money laundering, gambling and phishing scams, making cryptocurrency crimes a constraint for the industry and regulation. 
% In 2018, a statistical report published by the Russian cyber security company named Kaspersky Lab showed that Ethereum's ETH is the most popular digital asset for criminals, and the loss caused by illegal activities on the decentralized applications (DApps) has reached 900 million dollars.
Therefore, financial market security has become an important issue in the blockchain ecosystem, and it is of great significance to study security technology for public blockchain in application scenarios such as risk assessment and market supervision.

The biggest challenge for security supervision on blockchain is anonymity, which means that the identity information of each account need not to be disclosed or verified, resulting in the difficulty in mining privacy of account holders.
On the public blockchain, pseudonymous accounts created by criminals can protect their identity information to a certain extent and shelter their illegal activities from supervision.
On the other hand, the openness and transparency of blockchain provides access to all transaction information, which also creates conditions for the de-anonymization of accounts.
Recently, several related work has focused on using the public information available on blockchain platforms to analyze the behavior patterns of accounts and mine the identity information behind them, thus deriving a de-anonymization task -- account identity inference.
This task aims to infer the possible identity of accounts in blockchain, such as exchanges, phishing accounts, miners, Ponzi schemes, by using the public information about transaction, contract calls, etc.
Existing methods of account identity inference mainly concentrate on manual feature engineering, graph modeling and mining, which are effective but suffer from several shortcomings and challenges.
First, manual feature engineering relies on the prior knowledge of the designers, is not capable of capturing the underlying information in blockchain data, such as transaction patterns, resulting in low feature utilization and unsound expressiveness.
In addition, manual features have weak universality across different blockchain platforms due to the technical differences among them. 
For example, Ethereum has features associated with smart contract invocation that are not present in Bitcoin, which greatly limits the reusability of manually features.
Second, the scale of transaction graph in terms of the number of accounts and transactions is huge, resulting in high memory and time consumption when applying graph embedding algorithms based on random walks or graph neural network (GNN) to large-scale interaction graph.
Meanwhile, the growing number of transactions on the blockchain means the frequent update in the interaction graph in terms of nodes and edges, which is not conducive to whole-graph learning.
Lastly, the annotated data of account identity published by third-party sites is relatively scarce, resulting in a poor generalization of supervised models.

In our work, we present an end-to-end joint learning framework for account identity inference on \textbf{Block}chain with \textbf{G}raph \textbf{C}ontrast (BlockGC). 
Specifically, we first collect and collate large amounts of data involving transaction, contract invocation and public annotation data of account identity from the Ethereum and other third-party platforms.
Then we construct an account interaction graph using raw blockchain data.
Based on the assumption that the behavior patterns of different accounts are hidden in their neighborhood interaction subgraphs, we then extract neighborhood subgraphs of accounts from the interaction graph and pack them as batches as the input of model.
There exist few account identity labels in the existing blockchain data, which may lead to poor generalization of model during supervised learning.
So we introduce graph self-supervised learning for account identity inference. 
To be specific, several graph data augmentation strategies are firstly used to generate multiple weak annotation views of account subgraphs, and then a subgraph contrast process is used to learn the correlation between account subgraphs in different views.
Finally we jointly train the subgraph contrast and account classification task to realize the end-to-end account identity inference.

\section{Related Work}
Account identity inference in blockchain has received more and more attention with the purpose of detecting abnormal and illegal accounts. 
Previous methods concentrate on manual feature engineering.
Toyoda et al.~\cite{toyoda2018multi} extracted seven statistical features such as the rate of bitcoin coinbase transactions to infer account identity.
Lin et al.~\cite{lin2019evaluation} enriched the amount of features greatly and analyze the importance of each manual feature.
Bartoletti et al.~\cite{bartoletti2018data} designed the Gini coefficient and the characteristics of possible abnormal behavior patterns to infer the accounts of the Ponzi scheme in the transaction network.
Some emerging public blockchains contain smart contracts, providing new features. 
Huang et al.\cite{huang2020understanding} considered the calling information of smart contract to expand the feature space, and realized the identification of bot accounts in EOSIO.

Detecting abnormal users and learning their behavior patterns through transaction graph can be regarded as a node classification task in a graph perspective.
Therefore, the account identity inference method based on graph modeling emerged, providing convenience for analyzing complex transaction data.
Li et al.\cite{li2020identifying} considered topological features of accounts and found the difference in the topology between the Ponzi scheme accounts and the ordinary accounts.
Yuan et al.\cite{yuan2020detecting} applied DeepWalk\cite{perozzi2014deepwalk} and Node2vec\cite{grover2016node2vec} to learn account features in transaction graph.
Yuan et al.\cite{yuan2020phishing} extracted the subgraphs for each target account and embedded their transaction topology into feature vector via Graph2Vec\cite{narayanan2017graph2vec}.

\section{Account Interaction Graph Model}
\subsection{Ethereum and Block Data}
In this paper, we mainly focus on accounts in Ethereum which is the second largest blockchain platform after Bitcoin.
Ethereum allows users to conduct more complex transactions based on \textit{smart contracts}, which are applications that run on ethereum virtual machines. 
An \textit{account} in Ethereum is an entity that owns ether, and can be divided into two categories: external accounts (EOA) and contract accounts (CA). 
The EOA is controlled by user who owns the private key corresponding to the account, and can initiate transactions. 
The CA is controlled by code of smart contract, which cannot initiate transactions actively and can only be executed according to the pre-written code of smart contract after being triggered.
In Ethereum, there are usually two types of interactions: \textit{transaction} and \textit{contract call}.
The transaction must be initiated by EOA, and the receiver can be EOA or CA. 
Contract call refers to the process of triggering the smart contract code in CA to perform many different operations when an EOA initiates a transaction to the CA.

The Ethereum blockchain is a succession of blocks, each block contains a set of transactions $T$ and contract calls $C$. 
The raw block data of Ethereum is structured and provides a wealth of information, as listed in Appendix~\ref{tb: raw-data}.

\begin{figure}[htp]
\centering
  \includegraphics[width=\linewidth]{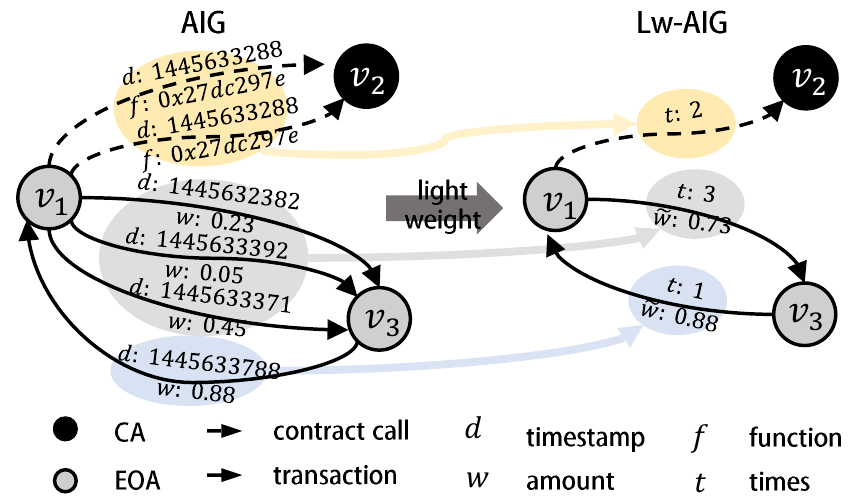}
  \caption{Account interaction graph and lightweight version.}
  \label{fig: AIG}
\end{figure}

\subsection{Account Interaction Graph}
The raw block data is informative and provides the details of transactions and contract calls, by which we can construct a Account Interaction Graph (AIG), as defined below.
\begin{definition}
	(\emph{Account Interaction Graph}): An Account Interaction Graph (AIG) is a directed and weighted multigraph $G = \left(V_{E O A}, V_{C A}, E_{T}, E_{C}, Y\right)$, where $V_{E O A}$ and $V_{C A}$ are the set of EOA and CA respectively, $E_{T}=\{e_{ij}^{T:d}=(v_{i}^{E O A}, v_{j}^{E O A}, d, w)\}$ is the directed edge set constructed from transaction information, and $E_{C}=\{e_{ij}^{C:d}=(v_{i}^{E O A+C A}, v_{j}^{C A}, d, f)\}$ is the directed edge set constructed from contract call information. 
	The three edge attributes $d$, $w$, $f$ represent the fields \textit{timestamp}, \textit{value} and \textit{callingFunction} respectively. 
	The AIG is partially labeled, i.e., a few accounts have identity labels and can compose the annotation node set $V_Y=\{(v_i, y_i) \mid v_i \in V_{EOA}\cup V_{CA}, y_i\in Y \}$.
\end{definition}
The original AIG is a multigraph which has a dense connection, as shown in Figure~\ref{fig: AIG}. 
We further ligntweight the AIG by edge coarsening and feature pruning to reduce the task complexity.
\begin{definition}
	(\emph{Lightweight Account Interaction Graph}): An Lightweight AIG (Lw-AIG) is a directed and weighted graph $G = (V_{E O A}, V_{C A}, \tilde{E} _{T}, \tilde{E} _{C}, Y)$, where $\tilde{E} _{T}=\{e_{ij}^{T}=(v_{i}^{E O A}, v_{j}^{E O A}, t, \tilde{w} )\}$ and $E_{C}=\{e_{ij}^{C}=(v_{i}^{E O A+C A}, v_{j}^{C A}, t)\}$.
	The edge attribute $t$ denotes the number of directed interactions from $v_i$ to $v_j$, and the edge attribute $\tilde{w}$ denotes total amounts of transaction from $v_i$ to $v_j$.
	
\end{definition}

Notably, the edge coarsening operation will merge all edges between any two nodes $v_i$, $v_j$ and generate an edge weight $t=|\{e_{ij}^{*:d}\}|$.
And the feature pruning will take effect with the edge coarsening, removing the two edge attribute of timestamp $d$ and function name $f$.

\begin{figure}[htp]
	\centering
  \includegraphics[width=\linewidth]{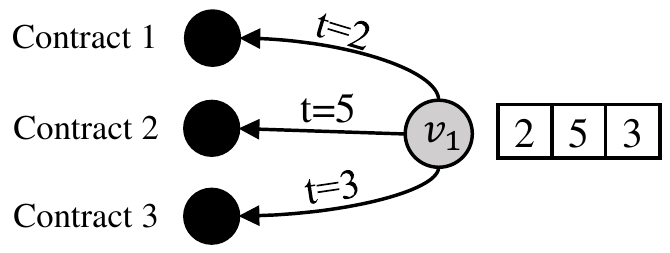}
  \caption{Illustration of constructing account attribute feature for contract call.}
  \label{fig: feature-cc}
\end{figure}

\subsection{Contract Call Feature}
The behavioral characteristics of an account are not only related to its transaction object, transaction amount and frequency, but also related to the smart contracts it calls. 
Different types of accounts have a certain correlation to different smart contracts. 
Therefore, the contract call situation of accounts can be treated as an account attribute feature, as illustrated in Figure~\ref{fig: feature-cc}.
We can construct a feature matrix $X\in\mathbb{R}^{n\times F}$ for the EOA in AIG to represent the preference for contract calls, and $F$ is the number of types of contracts.

% \begin{figure*}[htp]
% 	\centering
%   \includegraphics[width=\textwidth]{image/framework/sample-2.pdf}
%   % \setlength{\abovecaptionskip}{-7pt}ÔÚ¡¤
%   \caption{Example of sampling 2-hop subgraph.}
%   \label{fig: sample-2}
% \end{figure*}

\section{Methodology}
\begin{figure}[htp]
	\centering
  \includegraphics[width=\linewidth]{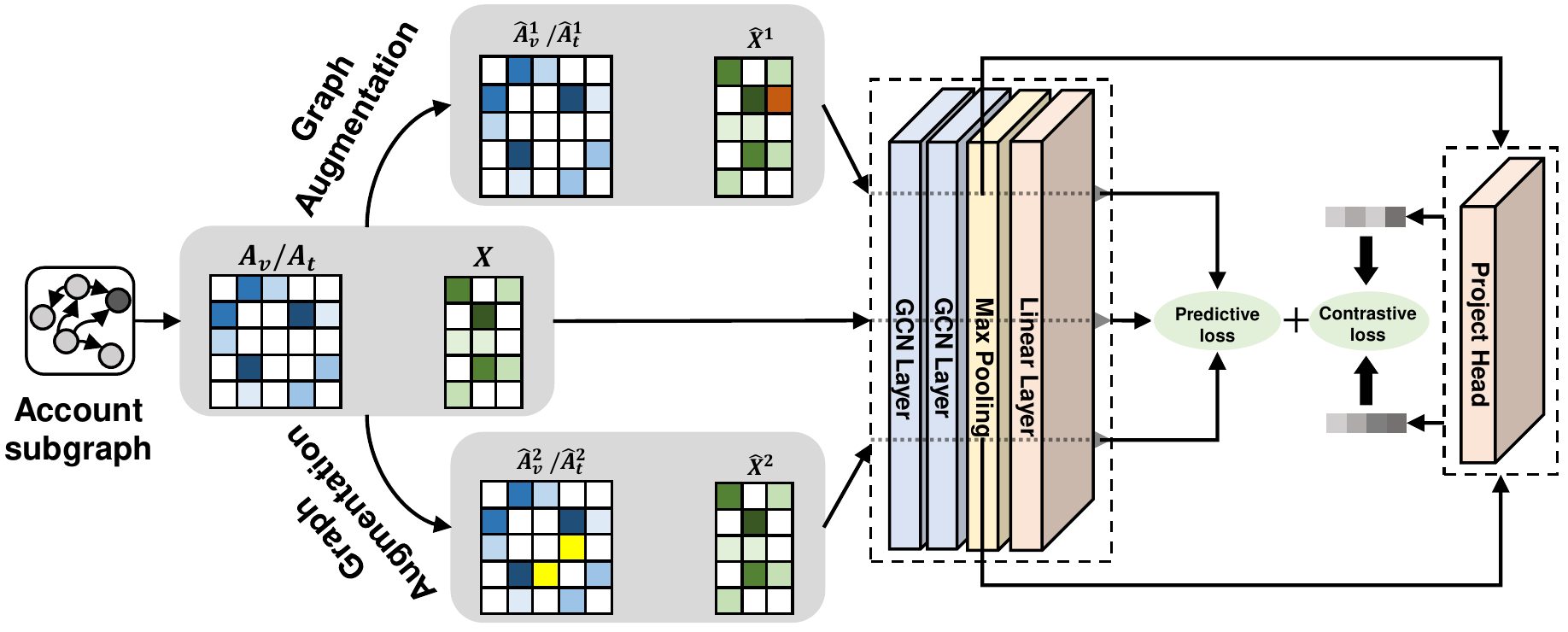}
  \caption{The architecture of BlockGC.}
  \label{fig: model}
\end{figure}

In this section, we give the details of the proposed framework BlockGC, as schematically depicted in Figure~\ref{fig: model}. 
Our framework is mainy composed of the following components: 
(1) a subgraph extractor which captures the subgraphs centered on target accounts from the Lw-AIG topology; 
(2) a subgraph augmentor which generates a series of variant graph views using various transformations on attributes and structure of subgraphs; 
(3) a GNN model which jointly trains the subgraph contrast and account classification tasks.
% (4) a subgraph-level contrast maximizes the consistency between two augmented views of the same subgraph. 
Next, we describe the details of each component.

\subsection{Subgraph Sampling}
The raw data contained tens of millions of blocks, making AIG a large-scale graph.
Some hub nodes such as exchange accounts have a large amount of transaction flows, which will affect both the computation efficiency and training effect.
Even though the lightweight process greatly sparses the connections of AIG, it does not reduce the number of nodes.
On the other hand, existing account identity inference methods based on random walk or graph embedding generally accept the whole graph as input and rely on full-graph training, which restricts their scalability on large-scale graphs for node representation learning.
In this work, we consider subgraph sampling strategies~\cite{chiang2019cluster, Zeng2020GraphSAINT} which are commonly used for large-scale graph computation to improve the scalability of our framework.

\begin{figure}[htp]
	\centering
  \includegraphics[width=\linewidth]{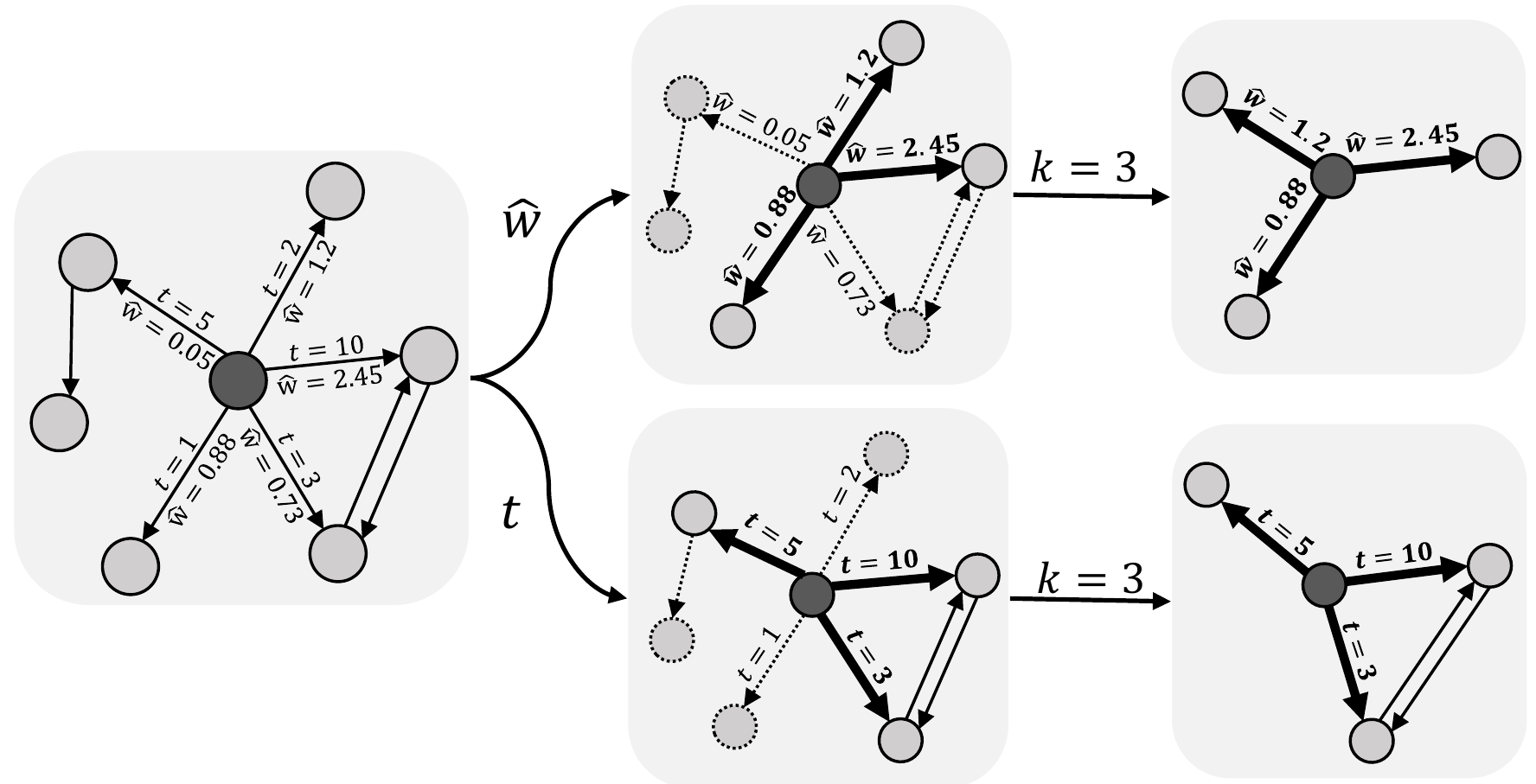}
  \caption{Illustration of Topk sampling according to different edge attributes.}
  \label{fig: sample-1}
\end{figure}

The Lw-AIG has two edge attributes: the number of interactions $t$ and the total amount of transactions $\tilde{w}$. 
Various subgraphs can be obtained by sampling according to different attributes.
We use Topk sampling to obatin the $h$-hop subgraph for each account, and convert the full-graph training to subgraph mini-batch training.
Specifically, for a target account node $v_i$, we sample 1-hop neighbors accoring to edge attributes with top-$k$ largest $t$ or $\tilde{w}$, and again sample the 1-hop neighbors for each nodes sampled at the previous hop, and recursive ones in the downstream hops.
After $h$ iterations, we obtain the node set $V_i$ sampled from Lw-AIG, and the subgraph $g_i$ of account $v_i$ can be induced by $V_i$ from the Lw-AIG.
Figure~\ref{fig: sample-1} illustrates the process of subgraph sampling according to difference edge attributes.
For a set of target accounts $V_Y$, we get the subgraph for each target account in $V_Y$ and all the subgraphs form a set: $D = \{(g_{i}, y_i)\mid v_i \in V_Y, y_i\in Y \}$.

\subsection{Subgraph Contrastive Learning}
BlockGC utilizes subgraph-level contrastive learning to train a GNN encoder $f$ which is capable of characterizing the interaction pattern of accounts.
In our contrastive learning framework, for each account subgraph $g_{i}$, its two correlated views $\hat{g}_i^1$ and $\hat{g}_i^2$ are generated by undergoing two augmentation operators $t_1$ and $t_2$, where $\hat{g}_i^1 = t_1(g_{i})$ and $\hat{g}_i^2 = t_2(g_{i})$.
The correlated augmented views are fed into a GNN encoder $f_\theta$ with max pooling layer, producing the whole subgraph representations $\mathbf{h}_i^1$ and $\mathbf{h}_i^2$, which are then mapped into an embedding space via a projection head $f_\phi $, yielding $\mathbf{z}_i^1$ and $\mathbf{z}_i^2$.
Note that $\theta$ and $\phi$ are the parameters of graph encoder and projection head respectively.
The representation of an account subgraph, $\mathbf{z}$, is treated as the representation of its central target account.
Finally, the goal of subgraph-level contrast is to maximize the consistency between two correlated augmented views of subgraphs in the embedding space via Eq.~(\ref{eq:loss-all}):
\begin{equation}\label{eq:loss-all}
  \mathcal{L}_{self} = \frac{1}{n} \sum_{i=1}^n  \mathcal{L}_i ,
\end{equation}
where $n$ is the number of subgraphs in a batch (i.e., batch size). The loss for each subgraph $\mathcal{L}_i$ can be computed as:
\begin{equation}\label{eq:loss-one}
  \mathcal{L}_i = -\log \frac{e^{\mathrm{s}\left(\mathbf{z}_{i}^{1}, \mathbf{z}_{i}^{2}\right) / \tau}}{\sum_{j=1, j \neq i}^{n} e^{\mathrm{s}\left(\mathbf{z}_{i}^{1}, \mathbf{z}_{j}^{2}\right) / \tau}},
\end{equation}
where $\mathrm{s}(\cdot, \cdot)$ is the cosine similarity function having $\mathrm{s}(\mathbf{z}_{i}^{1}, \mathbf{z}_{i}^{2}) = {\mathbf{z}_{i}^{1}}^\top\cdot \mathbf{z}_{i}^{2} / \| \mathbf{z}_{i}^{1}\| \| \mathbf{z}_{i}^{2}\|$, and $\tau$ is the temperature parameter.
The two correlated views $\mathbf{z}_i^1$ and $\mathbf{z}_i^2$ of account subgraph $g_{i}$ are treated as positive pair while the rest view pairs in the batch are treated as negative pairs.
The objective aims to maximize the consistency of positive pairs as opposed to negative ones.
% Note that here we use an asymmetrtic and simplified loss compared to the SimCLR loss~\cite{chen2020simple}, i.e., we generate negative pairs by only treating view 1 ($\mathbf{z}_i^1$) as the anchor and contrasting with view 2 ($\mathbf{z}_j^2$) of all other subgraphs, as shown in Eq.~(\ref{eq:loss-one}).

\subsection{Graph Augmentation}
Contrastive learning relies heavily on well-designed data augmentation strategies for view generation.
So far, commonly used graph augmentation techniques concentrate on structure-level and attribute-level augmentation.
In this paper, we use two existing augmentation methods, \emph{Node Dropping} and \emph{Feature Masking}.

\subsubsection{Feature Masking (FM)} 
It was proposed in~\cite{zhu2020deep}.
During feature masking, each dimension of node attributes has a certain probability $p$ to be set as zero.

\subsubsection{Node Dropping (ND)}
It was proposed in~\cite{you2020graph}.
During node dropping, each node has a certain probability $p$ to be dropped from $g_i$. 

During graph augmentation, we generate two augmented views $\hat{g}_i^1$, $\hat{g}_i^2$ for each target account subgraph $g_i$, and assign the label of target account to them as a pseudo label:
\begin{equation}
	\begin{array}{l}
		D_{aug1}=\left\{\left(\hat{g}_{i}^{1}, y_{i}\right) \mid v_{i} \in V_{Y}, y_{i} \in Y\right\} ,\\
		D_{aug2}=\left\{\left(\hat{g}_{i}^{2}, y_{i}\right) \mid v_{i} \in V_{Y}, y_{i} \in Y\right\}.
		\end{array}
\end{equation}

\subsection{Model Training}
We achieve account identity inference by classifying account subgraphs which consist of target account subgraphs and their corresponding augmented views, yielding a classification loss:
\begin{equation}
	\mathcal{L}_{pred} =  \frac{1}{3} (\mathcal{L}_{D}+\mathcal{L}_{D_{aug1}}+\mathcal{L}_{D_{aug2}}),
\end{equation}
where $\mathcal{L}_D$, $\mathcal{L}_{D_{aug1}}$ and $\mathcal{L}_{D_{aug2}}$ are the cross entropy loss computed by $D$, $D_{aug1}$ and $D_{aug2}$, respectively.

The subgraph contrast is treated as pretext task, and the encoder in BlockGC is jointly trained with the pretext and subgraph classification tasks.
The loss function consists of both the self-supervised and classification task loss functions, as formularized below:
\begin{equation}
	\mathcal{L} = \mathcal{L}_{pred}   + \lambda \cdot \mathcal{L}_{self}.
\end{equation}
where $\lambda$ is a trade-off hyper-parameter controls the contribution of self-supervision term.

\section{Experiments}
\subsection{Dataset Description}
We intercepte the first 10 million block data (the time interval is between ``2015-07-03'' to ``2020-05-04'') from the Xblock website\footnote{http://xblock.pro/}.
Within this time interval, we can extract in total 309010831 transactions and 175,351,541 contract calls, involving 90,193,755 EOA and 16,221,914 CA.
Account labels are obtained from Ethereum blockchain browser\footnote{https://etherscan.io/}, including 193 \emph{Exchange}, 73 \emph{ICO-wallets}, 65 \emph{Mining} and 2,535 \emph{Phish-hack}.
% \begin{itemize}
% 	\item \emph{Exchange}: 193 accounts.
% 	\item \emph{ICO-wallets}: 73 accounts.
% 	\item \emph{Mining}: 65 accounts.
% 	\item \emph{Phish-hack}: 2535 accounts.
% \end{itemize}
These four types of accounts are common in cryptocurrency networks, and have received widespread attention. 
It is of sufficient practical significance to inference whether an account belongs to these types, especially to identify phishing and hacker accounts.

\begin{figure*}[htp]
	\centering
  \includegraphics[width=\textwidth]{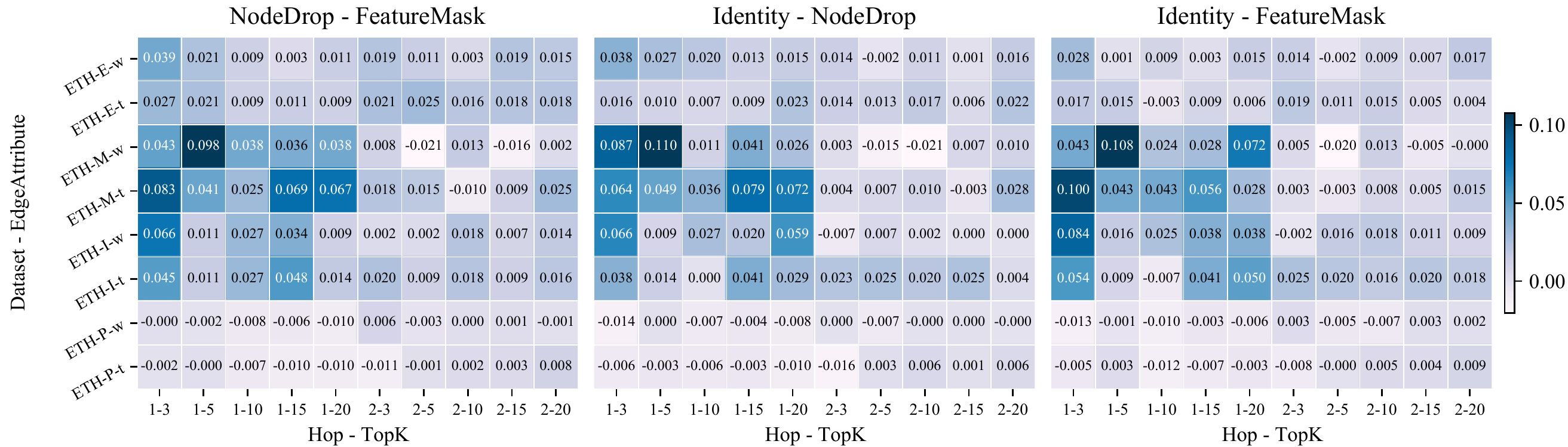}
  \caption{Performance gain (absolute improvement in F1 score) under different subgraph scales and data augmentation combinations, compared to I$^2$BGNN which stands for a no-augmentation baseline.}
  \label{fig: hotmap}
\end{figure*}

\subsection{Experimental Settings}
\subsubsection{Comparison Methods}
To illustrate the effectiveness, we compare the model with the following methods: manual features (see Appendix~B), DeepWalk~\cite{perozzi2014deepwalk}, Node2Vec~\cite{grover2016node2vec}, Struc2vec~\cite{figueiredo2017struc2vec}, I$^2$BGNN~\cite{shen2021identity}.

\subsubsection{Parameter Configuration}
For BlockGC, we set the subgraph hop $h$ to 2 and sample $k=20$ neighbors per hop.
We apply 2 layers of GCNs with output dimensions both equal to 128, and set \{$\lambda$, eopchs, batch size, dropout\} to be \{0.2, 200, 150, 0.3\}, respectively.
We repeat 3-fold cross validation for 10 times and report the average F1-score.
More details of parameter configurations for baselines are available in Appendix~D.

\begin{table}[htp]
	\renewcommand\arraystretch{1}
	\centering
	\caption{Results of identity inference for four types of accounts in Ethereum.}
	\label{tb: result}
	\resizebox{\linewidth}{!}{%
	\begin{tabular}{llccccc} 
		\toprule
		\multicolumn{2}{c}{Method}                        & Sampling           & ETH-E            & ETH-I            & ETH-M            & ETH-P  \\ 
		\midrule                      
		\multicolumn{2}{c}{Manual Feature}                 &    --             & 87.34            & 70.29            & 77.15            & 78.60  \\ 
		\midrule                      
		\multicolumn{2}{c}{\multirow{2}{*}{DeepWalk}}     & $\tilde{w}$        & 72.87            & 57.82            & 56.65            & 88.73  \\
		\multicolumn{2}{c}{}                              & $t$                & 77.61            & 61.88            & 57.46            & 89.33  \\
		\multicolumn{2}{c}{\multirow{2}{*}{Node2Vec}}     & $\tilde{w}$        & 80.79            & 66.21            & 69.51            & 89.30  \\
		\multicolumn{2}{c}{}                              & $t$                & 80.45            & 75.28            & 78.84            & 92.24  \\
		\multicolumn{2}{c}{\multirow{2}{*}{Struc2Vec}}    & $\tilde{w}$        & 72.01            & 56.24            & 74.36            & 64.27  \\
		\multicolumn{2}{c}{}                              & $t$                & 75.11            & 57.60            & 66.41            & 63.63  \\ 
		\midrule                               
		\multicolumn{2}{c}{\multirow{2}{*}{I$^2$BGNN}}    & $\tilde{w}$        & 87.94            & \underline{90.48}& 89.48            & \underline{96.07}  \\
		\multicolumn{2}{c}{}                              & $t$                & \underline{89.06}& 89.80            & \underline{90.27}& 95.43  \\ 
		\midrule                      
		\multirow{6}{*}{BlockGC} & \multirow{2}{*}{ND-FM} & $\tilde{w}$        & 89.49            & \textbf{91.84}   & 89.72            & 95.95   \\
								 &                        & $t$                & 90.87            & 91.28            & 92.76            & 96.21  \\
								 \cmidrule{2-7}                        
								 & \multirow{2}{*}{ND}    & $\tilde{w}$        & 89.66            & 91.38            & 89.46            & 96.03  \\
								 &                        & $t$                & 89.49            & 91.61            & 91.78            & 96.05  \\
								 \cmidrule{2-7}                      
								 & \multirow{2}{*}{FM}    & $\tilde{w}$        & 89.58            & 90.48            & 90.51            & 96.28  \\
								 &                        & $t$                & \textbf{91.30}   & 90.25            & \textbf{93.07}   & \textbf{96.29}  \\
		\bottomrule
	\end{tabular}}
  \end{table}

\subsection{Results and Analysis}
Table~\ref{tb: result} reports the performance comparison between BlockGC and baselines, from which we can observe that BlockGC significantly outperforms other methods across four types of accounts.
For instance, we achieve 91.30\%, 91.84\%, 93.07\% and 96.29\% F1-score, respectively, which are 2.52\%, 1.5\%, 3.1\% and 0.23\% relative improvement over previous state-of-the-art.

We further investigate the impact of two data augmentation techniques and the subgraph sampling strategy in our framework on Ethereum datasets, as illustrated in Figure~\ref{fig: hotmap}.
And we list the following \textbf{Obs}ervations.
More results will be discussed in Appendix.

\subsubsection{Obs. 1. Data augmentation is crucial, and feature masking benefits more.}
Composing an original and its FM view can benefit the inference task and achieve the best performance in 3 out of 4 experiments, judging from Table~\ref{tb: result}.
Since that the initial features of accounts in Lw-AIG is constructed from contract call information, which can reflect the behavioral preferences of accounts.
Masking such features can facilitate model learning the correlation between different behavioral views and help in identity inference. 
Such result is also consistent with ~\cite{you2020graph}, that is, feature masking performs better on dense graphs, like our AIG.

\subsubsection{Obs. 2. Smaller subgraphs contain more critical identity-related pattern information.}
It is generally better to infer from 1-hop subgraphs than from 2-hop ones, judging from Figure~\ref{fig: hotmap}.
Since AIG is a dense graph, the 2-hop subgraphs are large-scale, resulting in an over-smoothing in GNN. 
Moreover, the anonymity of blockchain also ensures that accounts in AIG only have partial knowledge of the 1-hop neighborhood accounts, resulting in a failure of message propagation from 2-hop neighbors.

\subsubsection{Obs. 3. Contrastive self-supervision facilitates the learning of account features.}
By contrastive learning, it seems that accounts of different classes on BlockGC with 2-dimensional embedding are more separated as compared to the embedding of I$^2$BGNN.

\begin{figure}[htp]
	\centering
  \includegraphics[width=0.8\linewidth]{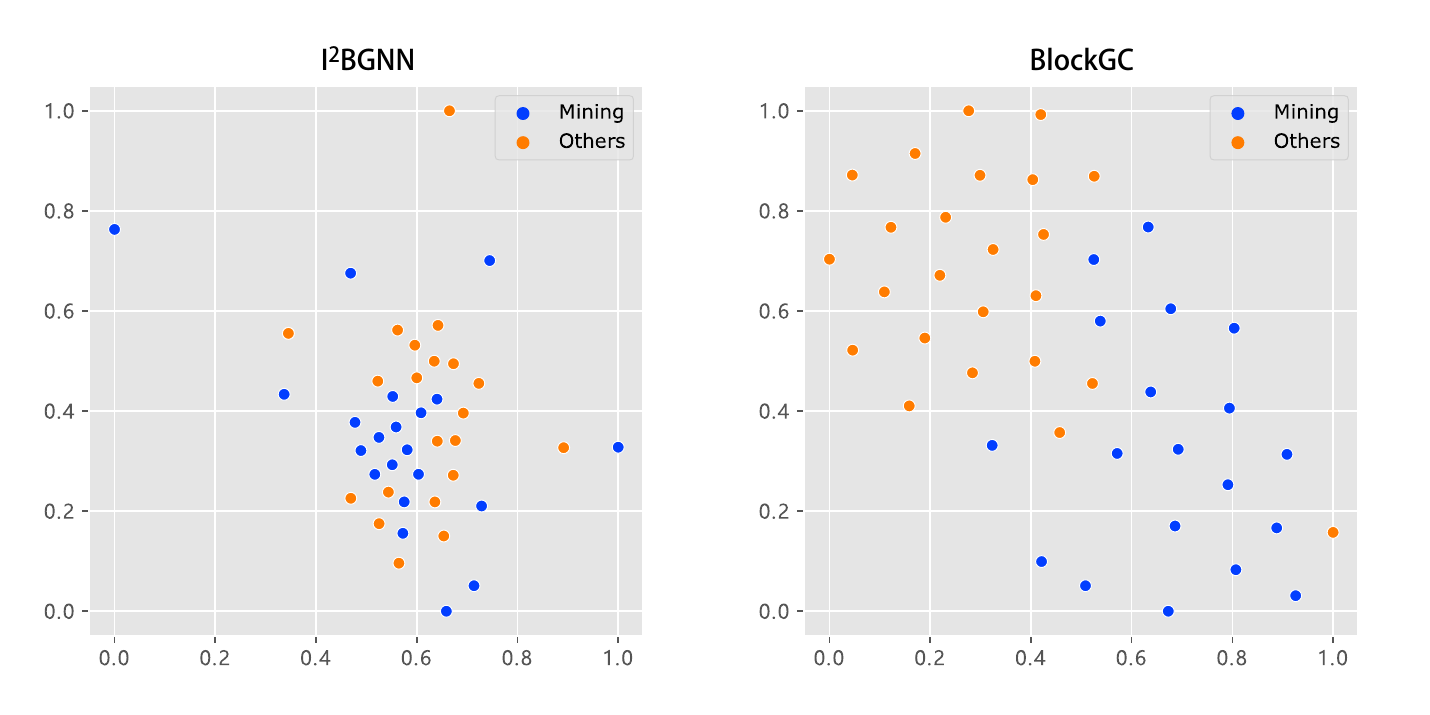}
  \caption{The t-SNE visualization of the embedding features obtained by I$^2$BGNN and our BlockGC.}
  \label{fig: tsne}
\end{figure}

  \section{Conclusion}
  In this paper, we construct an account interaction graph using raw blockchain data collected from Ethereum in a graph perspective, and proposes a joint learning framework for account identity inference on blockchain with graph contrast, which integrate scalability, generalization and end-to-end architecture together.
  Extensive experiments on Ethereum datasets show the superiority of our method.
% Use \bibliography{yourbibfile} instead or the References section will not appear in your paper
\bibliography{aaai22}

\clearpage
% \newpage

\appendix

\section{A. Ethereum Raw Block Data Information}
The raw block data collected from Ethereum is structured and provides a wealth of information, as listed in Table~\ref{tb: raw-data}.
\begin{table}[htp]
	\renewcommand\arraystretch{1}
	\centering
	\caption{Ethereum raw block data information.}
	\label{tb: raw-data}
	\resizebox{\linewidth}{!}{%
	\begin{tabular}{lcr} 
		\toprule
		\begin{tabular}[c]{@{}l@{}}Data~\\Field\end{tabular} & \begin{tabular}[c]{@{}l@{}}Custom \\Symbol\end{tabular} & Definition     \\
		\midrule
		blockNumber     & -               & The ID of block where the transcation is located                             \\
		timestamp       & $d$             & The timestamp of the transaction                                       \\
		from            & $v$             & The account that initiates the transaction                             \\
		to              & $v$             & The account that receives the transaction                              \\
		fromIsContract  & -               & Whether the transaction is initiated from a CA\\
		toIsContract    & -               & Whether the transaction is received by a CA  \\
		callingFunction & $f$             & The name of function called if there is a contract call                 \\
		value           & $w$             & The transaction amount                                                 \\
		\bottomrule
	\end{tabular}}
  \end{table}

\section{B. Manual Feature Details} \label{app: manual}
Manual feature engineering is the most common and simplest way for account identity inference. 
According to the characteristics of raw transaction data and prior knowledge, we design 16 manual features for Ethereum accounts, as shown in Table~\ref{tb: manual}.
\begin{table}[htp]
\centering
\renewcommand\arraystretch{1}
\caption{Statistics of the average of manual feature for various accounts in Ethereum.}
\label{tb: manual}
\resizebox{\linewidth}{!}{%
\begin{tabular}{lccccc} 
\toprule
Manual Features            & Phish-Hack & Exchange   & Mining    & ICO-Wallets & Common  \\ 
\midrule
active\_days               & 76.94      & 703.35     & 595.98    & 547.84      & 14.49   \\
total\_received            & 110.84     & 1551629.77 & 5470.67   & 6642.11     & 245.38  \\
num\_received\_tx          & 27         & 88490.39   & 68.55     & 279.86      & 4.13    \\
inter\_acct\_received      & 23.29      & 29985.26   & 13.6      & 218.77      & 0.4     \\
total\_output              & 124.03     & 2309107.24 & 367339.66 & 33824.02    & 370.21  \\
num\_output\_tx            & 29.91      & 88285.11   & 818877.92 & 62.84       & 4.56    \\
inter\_acct\_output        & 8.18       & 46692.69   & 16825.85  & 37.53       & 1.48    \\
avg\_received              & 29.52      & 2002.55    & 49.79     & 854.35      & 7.06    \\
avg\_received\_day         & 11.66      & 1748.8     & 6.55      & 11.36       & 4.93    \\
avg\_received\_tx\_day     & 1.51       & 113.5      & 0.16      & 0.49        & 0.03    \\
avg\_output                & 30.58      & 1453.75    & 249.95    & 4399.06     & 9.46    \\
avg\_output\_day           & 14.66      & 2213.77    & 389.08    & 80.3        & 5.39    \\
avg\_output\_tx\_day       & 0.66       & 101.26     & 633.21    & 0.2         & 0.03    \\
times\_contract\_called      & 11.68      & 66682.01   & 61002.03  & 690.93      & 1.52    \\
times\_contract\_called\_day & 0.5        & 82.66      & 41.84     & 1.07        & 0.02    \\
num\_contract\_called    & 3.29       & 2031.6     & 1256.31   & 3.26        & 0.13    \\
\bottomrule
\end{tabular}
}
\end{table}

The features are described as follows:

\begin{itemize}
    \item \emph{active\_days}: the number of days the account has been active.
    \item \emph{total\_received}: the sum of ether received by the account.
    \item \emph{num\_received\_tx}: the number of transactions that the account has received ether.
    \item \emph{inter\_acct\_received}: the number of others interacted with the account in the transactions that the account has received ether.
    \item \emph{total\_output}: the value of ether spent by the account.
    \item \emph{num\_output\_tx}: the number of transactions that the account has spent ether.
    \item \emph{inter\_acct\_output}: the number of others interacted with the account in the transactions that the account has spent ether.
    \item \emph{avg\_received}: the average value of the ether received by the account.
    \item \emph{avg\_received\_day}: the average value of ether received by the account per day.
    \item \emph{avg\_received\_tx\_day}: the average number of transactions that the account has received ether per day.
    \item \emph{avg\_output}: the average value of the ether spent by the account.
    \item \emph{avg\_output\_day}: the average value of the ether spent by the account per day.
    \item \emph{avg\_output\_tx\_day}: the average number of transactions that the account has spent ether per day.
    \item \emph{times\_contract\_called}: the times the account has called the smart contract.
    \item \emph{times\_contract\_called\_day}: the times the account has called the smart contract per day.
    \item \emph{num\_contract\_called}: the number of contracts called by the account.
\end{itemize}

\section{C. Subgraph Details}
For each type of account identity label (Exchange, ICOwallets, Mining, PhishHack), we sample the subgraphs of all accounts with that label, as well as the same number of subgraphs of other types of accounts.
We perform Topk sampling for each labeled accounts according to the transaction amounts ($\tilde{w}$) or interaction times ($t$), yielding two types of subgraphs (``-w'', ``-t'') for each account.
We sample up to 20 neighbors for each account per hop, and the statistics of sampled subgraphs ($k=20$) are shown in Table~\ref{tb: subgraph}.

% the same number of positive and negative samples as the manual features-based method in Chapter 2 and the setting of trisector cross verification for data set. Second-order neighborhood subgraph is extracted for each sample, and each sub-graph samples 20 neighbor nodes at most. Descending sampling is performed according to the total amount of transactions between nodes ($V$) or the number of transactions between nodes ($T$). The features of each node are constructed from the situation about node's calling of the contract, EOSIO added three extra dimensions to describe the features of an account. Eventually, we get the Subgraph data set size, shown in table~\ref{tb: 4-1}, from the five data sets according to the two sampling strategy, including $\mid G\mid $ for the figure number, $\mid N\mid $ for the average number of nodes, $\mid E\mid $ for the average number of edges, $\mid F\mid $ for the feature dimension.

\begin{table}[htp]
\centering
\caption{Statistics of sampled subgraphs from Ethereum for accounts of different classes.
$|G|$ is the number of subgraphs, \textit{Avg.} $|N|$ and  \textit{Avg.} $|E|$ are the average number of nodes and edges in subgraphs respectively, and $|F|$ is the number of features of accounts in subgraphs.}
\label{tb: subgraph}
\resizebox{\linewidth}{!}{%
\renewcommand\arraystretch{1}

\begin{tabular}{lccccc} 
\toprule
Dataset              & Abbr.         & $|G|$      & \textit{Avg.} $|N|$   & \textit{Avg.} $|E|$   & $|F|$    \\ 
\midrule    
% EOSIO-w              &         & 2000                        & 260.19                 & 3351.74               & 1216                   \\
% EOSIO-t              &         & 2000                        & 206.73                 & 2442.59               & 1216                   \\
ETH-Exchange-w     & ETH-E-w        & 386                         & 88.46                  & 215.01                & 14885                  \\
ETH-Exchange-t     & ETH-E-t        & 386                         & 101.77                 & 278.59                & 14885                  \\
ETH-ICOwallets-w   & ETH-I-w        & 146                         & 81.79                  & 180.75                & 14885                  \\
ETH-ICOwallets-t   & ETH-I-t        & 146                         & 83.08                  & 198.77                & 14885                  \\
ETH-Mining-w       & ETH-M-w        & 130                         & 94.31                  & 225.84                & 14885                  \\
ETH-Mining-t       & ETH-M-t        & 130                         & 94.96                  & 228.45                & 14885                  \\
ETH-PhishHack-w    & ETH-P-w        & 5070                        & 59.23                  & 124.92                & 14885                  \\
ETH-PhishHack-t    & ETH-P-t        & 5070                        & 59.76                  & 130.62                & 14885                  \\
\bottomrule
\end{tabular}
}
\end{table}

% \section{C. Details of I$^2$BGNN}
% There are the details of proposed I$^2$BGNN for identity inference on blockchain. For graph classification, the pooling operations aggregate node representations from the final iteration to obtain the whole graph’s representation. By stacking the pooling layer and fullyconnected layer after 2-layer GCN, the basic graph classification model for identity inference can be constructed as follows:
% \begin{equation}
% \begin{aligned}
% \mathbf{H} & =\mathrm{ReLU}\left(\widehat{\mathbf{A}} \cdot \mathrm{ReLU}\left(\widehat{\mathbf{A}} \mathbf{X} \mathbf{W}^{(0)}\right) \cdot \mathbf{W}^{(1)}\right) \\
% \mathrm{logits} & =\mathrm{Softmax}\left(\mathrm{MaxPooling}(\mathbf{H}) \mathbf{W}^{(2)}+\mathbf{b}\right)
% \end{aligned}
% \end{equation}
% Note that we use the max pooling to obtain the whole graph’s representation. The model architecture of I$^2$BGNN is finished. In a transaction subgraph, each node represents an account and each directed edge represents transaction flow that contains information about transaction volume and frequency. For the input layer of GCN, we first initialize the node representations using their attributions in transaction subgraph.

\section{D. Details of Parameter Configuration}\label{app: para}
For manual feature engineering, we design 16 manual features, as described in Appendix~B.
For DeepWalks and Struc2Vec, we set the length of walks to 20, the number of walks to 40, and the context size to 3. 
For Node2Vec, we set the return parameter $p$ and in-out parameter $q$ to 0.25 and 0.4, respectively.
For all the above baselines , we implement account identity inference by using Logistic Regression classifier.

For I$^2$BGNN, we set the output dimensions of 2 hidden layers both to 128. 

For all the baseline based on graph embedding, we set the dimension of output embedding to 128.

\begin{figure*}[htp]
	\centering
  \includegraphics[width=\textwidth]{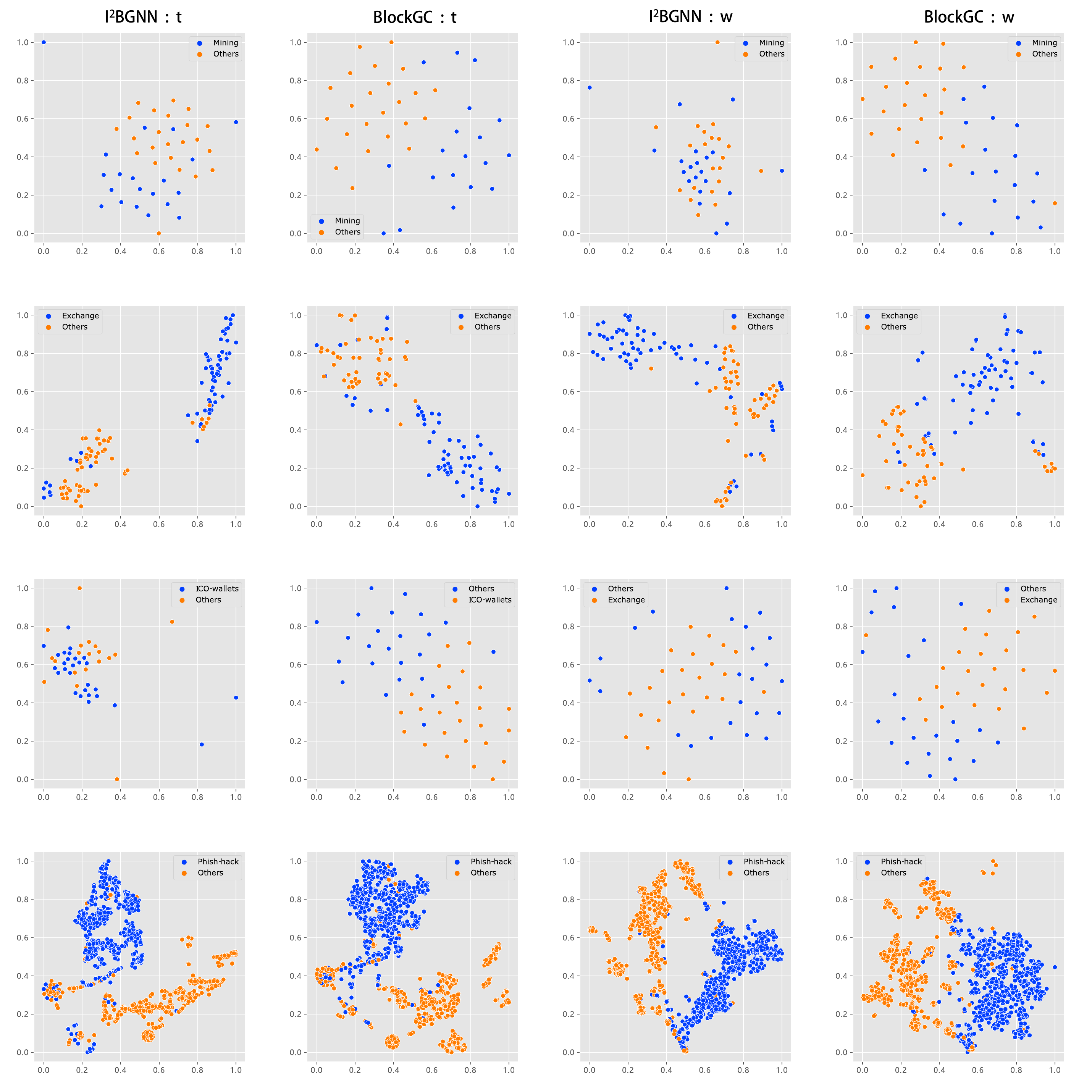}
  \caption{The t-SNE visualization of the embedding features obtained by I$^2$BGNN and our BlockGC in terms of four classes of accounts.}
  \label{fig: tsne-all}
\end{figure*}

\section{E. More Results and Analysis}

% \subsection{Subgraph Sample}
\subsubsection{Obs. 4. Subgraphs sampled with interaction times benefits more.}
As we can see from Table~\ref{tb: result}, infering with subgraphs sampled according to $t$ achieve better performance than those sampled with $\tilde{w}$ in 20 out of 28 experiments across all methods.
Since that interaction times $t$ reflects the frequency of transactions and contract invocations, while the transaction amounts $\tilde{w}$ refers only to the transaction process, and the former reflects more information for identity inference.

% \subsubsection{Obs. 5. The optimal sampling size and strategies varies by datasets, but they also have common patterns.} 
% Firstly, we find that picking the transaction amount is always better than the number of transactions on almost all types of account detection. 
% This is in line with our intuition: In the field of economics, it is the amount of money that is the universal calculator of relationships between people while the number of transactions may be fraudulent. 
% Our results show that the amount of transactions between blockchain accounts better reflects the information of the whole graph transaction characteristics than the number of transactions.
% According to \ref{Fig3}, the best sampling is specific to datasets, though our datasets are similar to each other. Mining account dataset achieved the top with 1-5 sampling. An improvement of 0.098, 0.110, and 0.108 was reached under the three data augmentations strategies of NF, IN, and IF, respectively. But on the 1-Exchange accounts, ICO-wallets and Phish-hack dataset, They all perform best with a 1-3 sampling strategy.

\subsubsection{Obs. 5. Contrastive self-supervision improve the generalization of model in learning account features.}
As we can see frim Figure~\ref{fig: tsne-all}, that accounts of different classes on BlockGC with 2-dimensional embedding tend to be more separated as compared to the embedding of I$^2$BGNN, showing clearer decision boundaries, which facilitate the classification of account identity labels.
Moreover, the feature embedding learned by BlockGC has a more uniform distribution, reflecting a more powerful generalization.

\end{document}